\documentclass[aps,showpacs,showkeys,nofootinbib,twocolumn]{revtex4}
\usepackage{epsfig,amssymb,bm,graphicx,color,amsmath,rotating}
\usepackage{xcolor}
\usepackage[percent]{overpic}
\usepackage{mwe,xcolor}
\usepackage{transparent}
\usepackage{framed}
\usepackage{slashed}
\newcommand*{\ee}{e^+e^-}

\begin{document} {\normalsize }
\interfootnotelinepenalty=10000

\title{Rise and fall of laser-intensity effects in spectrally resolved
Compton process}

\author{U. Hernandez Acosta${}^{1, 2, 3}$, A.~I.~Titov${}^4$, B.~K\"ampfer${}^{1, 2}$}
\affiliation{${}^1$Helmholtz-Zentrum  Dresden-Rossendorf, 01314 Dresden, Germany}
\affiliation{${}^2$Institut f\"ur Theoretische Physik, TU~Dresden, 01062 Dresden, Germany}
\affiliation{${}^3$Center for Advanced Systems Understanding, 
%Helmholtz-Zentrum Dresden-Rossendorf e.V. (HZDR)
Untermarkt 20, 02826 Görlitz, Germany}
\affiliation{${}^4$Bogoliubov Laboratory of Theoretical Physics, JINR, Dubna 141980, Russia}

\begin{abstract}
The spectrally resolved differential  cross section of Compton scattering,
$d \sigma / d \omega' \vert_{\omega' = const}$,
rises from small towards larger 
laser intensity parameter $\xi$, reaches a maximum, and falls towards the asymptotic 
strong-field region. Expressed by invariant quantities:
$d \sigma /du \vert_{u = const}$ rises from small towards larger 
values of $\xi$, reaches a maximum at 
$\xi_{max} = \frac49 {\cal K} u m^2 / k \cdot p$, 
${\cal K} = {\cal O} (1)$, and falls at $\xi > \xi_{max}$ like 
$\propto \xi^{-3/2} \exp \left (- \frac{2 u m^2}{3 \xi \, k \cdot p} \right )$
at $u \ge 1$.
[The quantity $u$ is the Ritus variable related to the light-front momentum-fraction 
$s = (1 + u)/u = k \cdot k' / k \cdot p$
of the emitted photon (four-momentum $k'$, frequency $\omega'$),
and $k \cdot p/m^2$ quantifies the invariant energy in the entrance channel
of electron (four-momentum $p$, mass $m$) and laser (four-wave vector $k$).]
Such a behavior of a differential observable is to be contrasted with the
laser intensity dependence of the total probability, 
$\lim_{\chi = \xi k \cdot p/m^2, \xi \to \infty} \mathbb{P} 
\propto \alpha \chi^{2/3} m^2 / k \cdot p$,
which is governed by the soft spectral part.

We combine the hard-photon yield from Compton with the seeded
Breit-Wheeler pair production in a folding model and obtain a rapidly
increasing $e^+ e^-$ pair number at $\xi \lesssim 4$.  
Laser bandwidth effects are quantified in the weak-field limit of the 
related trident pair production.

\end{abstract}


\date{\today}

\maketitle

\section{Introduction}

Quantum Electro-Dynamics (QED) as pillar of the standard model (SM)
of particle physics possesses a positive $\beta$ function \cite{Peskin:1995ev} which
makes the running coupling strength $\alpha (\mathfrak{s} )$ increasingly with
increasing energy/momentum scale $\mathfrak{s}$ \cite{Abbiendi:2005rx}.
In contrast, 
Quantum Chromo-Dynamics (QCD) as another SM pillar possesses a negative
$\beta$ function 
due to the non-Abelian gauge group \cite{Peskin:1995ev},
giving rise to the asymptotic freedom, 
$\lim_{\mathfrak{s} \to \infty} \alpha_{QCD} (\mathfrak{s}) \to 0$,
i.e.\ QCD has a truly perturbative limit. In contrast, 
$\lim_{\mathfrak{s} \to 0} \alpha (\mathfrak{s}) \to 1/137.0359895(61)$ is not such
a strict limit, nevertheless, QED predictions/calculations of some observables 
agree with measurements within 13 digits, see
\cite{Aoyama:2020ynm,Arapoglou:2019taq,QED_prec1}
for some examples.
The situation in QED becomes special when considering processes in external
(or background) fields: One can resort to the Furry (or bound-state) picture,
where the (tree-level) interactions of an elementary charge (e.g.\ an electron)
with the background are accounted for 
in all orders, and the interactions with the quantized photon field remains
perturbatively in powers of $\alpha$.
However, the  Ritus-Narozhny (RN) conjecture \cite{RN1,RN2,RN3,RN4,RN5} 
argues that the
effective coupling becomes $\alpha \chi^{2/3}$, meaning that the Furry
picture expansion beaks down at  $\alpha \chi^{2/3} > 1$ 
\cite{Mironov:2020gbi,Heinzl:2021mji,Edwards:2020npu,Fedotov:2016afw}
(for the definition of $\chi$ see below) 
and one enters a genuinely non-perturbative regime. 
The latter requires adequate calculation procedures, as the lattice regularized
approaches, which are standard since many years in QCD, e.g.\ in evaluations
of observables in the soft sector where $\alpha_{QCD} > 1$, 
cf.\ \cite{Brambilla:2014jmp}.
(In QED itself, an analog situation is meet in the Coulomb field of
nuclear systems with proton numbers $Z > Z_{crit} \approx 173$: 
if $\alpha Z_{crit} > 1$ the QED vacuum beak-down sets in; 
cf.\ \cite{Popov:2020xmd} for the actual status of that field).

With respect to increasing laser intensities the quest for the possible break-down
of the Furry picture expansion in line with the RN conjecture becomes, besides
its principal challenge, also of ``practical" interest, whether one can explore
experimentally this yet uncharted regime of QED. 
(For other configurations, e.g.\ beam-beam
interactions, cf.\ \cite{Yakimenko:2018kih}). A prerequisite would be to find
observables which display the typical dependence $\propto \alpha \chi^{2/3}$,
where we denote by $\alpha$ the above quoted fine-structure constant
at $\mathfrak{s} \to 0$.
In doing so we resort here to the lowest-order QED processes, that is nonlinear
Compton and nonlinear Breit-Wheeler. Both processes seem to be investigated
theoretically in depth in the past, 
however, enjoy currently repeated
re-considerations w.r.t.\ refinements 
\cite{DiPiazza:2020wxp,Seipt:2020diz,King:2020btz,Ilderton:2020dhs,Titov:2018bgy,Granz:2019sxb},
establishing approximation schemes 
\cite{Blackburn:2021rqm,Heinzl:2020ynb,DiPiazza:2018bfu,Blackburn:2018sfn,Ilderton:2018nws}
to be implemented in simulation codes  
\cite{Shvets:2018iar,Blackburn:2017dpn,Gonoskov:2014mda}
or to use them as building blocks in complex processes \cite{Torgrimsson:2020gws}, 
with the starting point at trident 
\cite{Torgrimsson:2020wlz,Dinu:2019wdw,Dinu:2019pau}.

Beginning with Compton scattering
($p$ and $p'$ are the $in$- and $out$-electron four-momenta,
$k$ the laser four-momentum, 
and $k'$ the $out$-photon's four-momentum, respectively) 
the relevant (Lorentz and gauge invariant) variables of the entrance channel are
\cite{DiPiazza:2011tq} \\
- (i) the classical intensity parameter of the laser:
$\xi =  \vert e \vert {\cal E} /(m \omega)$, here expressed
by quantities in the lab.: ${\cal E}$ - electric laser field strength, $\omega$ -
the central laser frequency; $- \vert e \vert$ and $m$ stand for the electron
charge and mass, respectively, and $e^2 / 4 \pi = \alpha$,\footnote{We employ natural units with $\hbar = c = 1$.}\\
- (ii) the available energy squared: $k \cdot p / m^2 = (\hat s /m^2 - 1)/2$
with $\hat s$ as Mandelstam variable,\\
- (iii) the quantum nonlinearity parameter: $\chi = \xi k \cdot p/m^2$.\\
The latter quantity is often considered as the crucial parameter since
in some limits it determines solely the probability of certain processes.
$\chi$ plays also a prominent role in the above mentioned
discussion of the RN conjecture, 
where the Furry picture expansion 
of QED is argued to break down for $\alpha \chi^{2/3} > 1$. 
References \cite{Ilderton:2019kqp,Podszus:2018hnz}
point out that the large-$\chi$ limits,
facilitated by either large $\xi$ (the high-intensity limit) or large $k \cdot p/m^2$
(the high-energy limit), are distinctively different, with implications for
approximation schemes in simulation codes.

\begin{figure}[tb!]
\includegraphics[width=0.99\columnwidth]{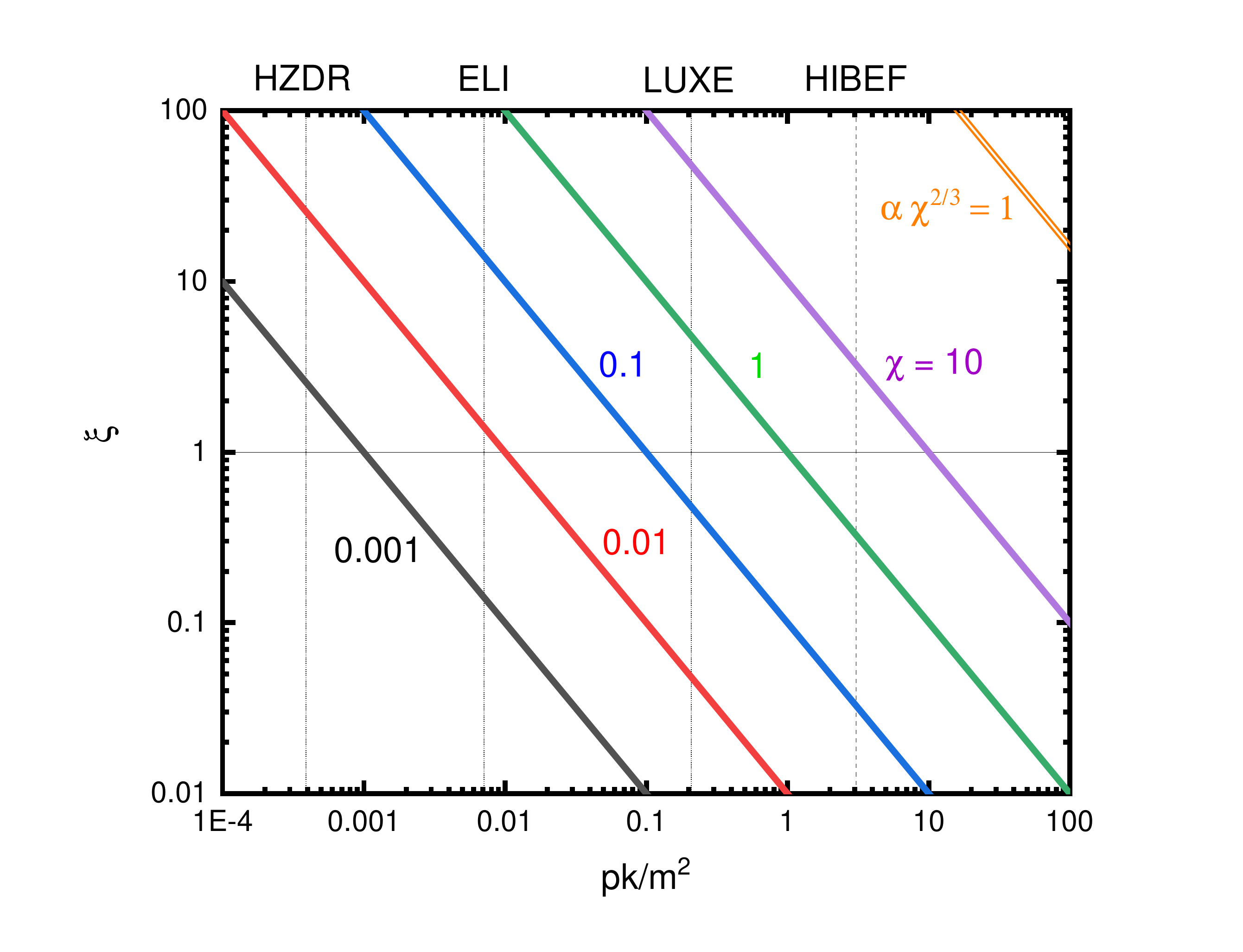}
\vspace*{-3mm}
\caption{Curves of $\chi = const$ over the $\xi$ vs.\ $k \cdot p/m^2$ plane
for $\chi = 0.001,$ 0.01, 0.1, 1 and 10.
The double-line depicts the curve $\alpha \chi^{2/3} = 1$.
Vertical thin lines are for maximum values of $k \cdot p/m^2$ in reach
at various electron accelerators 
(HZDR \cite{Jochmann:2013toa}: $E_e = 33$ MeV,
ELI \cite{Turcu:2016dxm}: $E_e = 600$ MeV, 
LUXE \cite{Abramowicz:2019gvx,Abramowicz:2021zja}: $E_e = 17.5$ GeV)
in combination with a high-intensity optical laser
(we use $\omega = 1.55$ eV as representative frequency).
The vertical dotted line indicates a possible combination of the
European XFEL ($\omega = 10$ keV) with a laser-accelerated electron
beam ($E_e = 10$ MeV) available in the high-energy density cave
of the HIBEF collaboration \cite{HIBEF}.
The horizontal delineation line $\xi = 1$ is thought to highlight the
onset of the strong-field region above.
\label{a0_pk_landscape}}
\end{figure}

Figure \ref{a0_pk_landscape} exhibits a few selected curves $\chi = const$
over the $\xi$ vs.\ $k \cdot p/m^2$ landscape to illustrate the current situation
w.r.t.\ facilities where laser beams and electron beams are (or can be) combined.
One has to add the options at E-320 at FACET-II/SLAC 
\cite{E_320,Meuren:2020nbw} and
electron beams which are laser-accelerated to GeV scales, e.g.\
\cite{Poder:2018ifi,Cole:2017zca,Keitel:2021gcq}.

Usually, $\xi = 1$ is said to mark the the onset of strong-field effects, and the
corresponding processes at $\xi > 1$ are termed by the attribute ``nonlinear".
In this respect, the parameters provided by LUXE 
\cite{Abramowicz:2019gvx,Abramowicz:2021zja}
and FACET-II \cite{E_320,Meuren:2020nbw} are interesting:
$\chi = {\cal O}(1)$ and above and $\xi > 1$ as well.\footnote{
Despite high intensities
at XFELs, e.g.\ $I \to 10^{22}$ W/cm${}^2$, the intensity parameter 
$\xi = \frac{7.5 {\rm eV}}{\omega} \sqrt{\frac{I}{10^{20} {\rm W/cm}^2}}$
\cite{DiPiazza:2011tq}
is small due to the high frequency, e.g.\ $\omega =$ 1 - 25 keV.}  
In the following we consider the LUXE kinematics (see Fig.~\ref{a0_pk_landscape}),
$k \cdot p/m^2 \approx \omega E_e (1 - \cos \Theta) = 0.2078$
in head-on collisions ($\cos \Theta = -1$). Our aim is to quantify a simple
observable as a function of $\xi$. To be specific, we select the invariant
differential cross section $d \sigma / du$, where $u$ is the
light-cone momentum-transfer of the $in$-electron to the $out$-photon,
related to light-front momentum-fraction of the $out$-photon
$s = \frac{u}{1 + u} = \frac{k \cdot k'}{ k \cdot p}$.
(The mapping $u \mapsto \omega'$ and 
$d \sigma / du \mapsto d \sigma / d \omega'$ is discussed in 
\cite{Acosta:2020agu,Kampfer:2020cbx}.)
To make the meaning of the Ritus variable $u$ more explicit let us
mention the relation
$u = \frac{e^{-\zeta} \nu' (1 - \cos \Theta')}{1 - 
e^{-\zeta} \nu' (1 - \cos \Theta')}$, where $\nu' \equiv \omega'/m$, and
the electron energy $E_e$ in lab.\ determines
the rapidity $\zeta$ via $E_e = m \cosh \zeta$ and $\Theta'$ denotes
the polar lab.\ angle of the $out$-photon.
We call $d \sigma / du$ a spectrally resolved observable. 

Our note is organized as follows. In section \ref{laser_models}
we briefly recall a few approximations of the laser beam.
Section \ref{Compton} is devoted to an analysis of the invariant
differential cross section $d \sigma/du$ and its dependence on the
laser intensity parameter $\xi$ in nonlinear Compton scattering.
That is, we are going up and down on the vertical dashed line 
with label LUXE in Fig.~\ref{a0_pk_landscape} around the point 
$\chi = 1$ or $\xi = 1$.
The discussion section \ref{discussion} (i) relates the cross section
to the probability and (ii) considers a folding model which uses the
hard-photon spectrum emerging from Compton scattering as seed
for subsequent Breit-Wheeler pair production; a brief discussion of (iii)
bandwidth effects relevant for sub-threshold trident pair production 
complements this section.
We conclude in section \ref{summary}.
The appendix \ref{App_A} recalls a few basic elements of 
the one-photon Compton process.

\section{Laser models}\label{laser_models}

In plane-wave approximation\footnote{Due to the high symmetry, the exact
solutions of the Dirac equation in a plane-wave background are in a comfortably compact form
enabling an easy processing in evaluations of matrix elements.
Without such a high symmetry, much more attempts are required
\cite{DiPiazza:2021rum,DiPiazza:2016tdf,Heinzl:2017zsr}.
For some useful parameterizations of laser beams, see
\cite{Karbstein:2020gzg,Blinne:2018nbd,Karbstein:2017jgh} and further citations therein.
The relevance of the Fourier-zero mode of (non-)unipolar planar fields 
is mentioned in \cite{Dinu:2012tj}.}
the laser (circular polarization) can be described by the four-potential
in axial gauge, $A = (0, \vec A)$, with
\begin{equation}\label{laser}
\vec A = f(\phi) \left(\vec a_x \cos \phi + \vec a_y \sin \phi \right)
\end{equation}
where $\vec a_x^2 = \vec a_y^2 = m^2\xi^2/e^2$;
the polarization vectors $\vec a_x$ and  $\vec a_y$ are mutually
orthogonal. We ignore a possible non-zero value
of the carrier envelope phase and focus on symmetric envelope functions
$f(\phi)$ w.r.t.\ the invariant phase $\phi = k \cdot x$. 
One may classify the such a model class as follows.\\
- 1) Laser pulses: $\lim_{\phi \to \pm \infty} f(\phi) = 0$,\\
- 2) Monochromatic beam: $f(\phi) = 1$,\\
- 3) Constant cross field (ccf): $\vec A = \phi \, \vec a_x $.\\
The probabilities for the constant cross field option 3) coincide with certain limits
of the plane-wave model 2) \cite{Ritus}; in \cite{Kampfer:2020cbx} 
they are related to the large-$\xi$ limit.
Item 1) could be divided into several further sub-classes, such as
1.1): finite support region of the pulse, i.e.\
$f(\vert \phi \vert > \phi_{pulse \, length}) = 0$, 
$\phi_{pulse \, length} < \infty$, and
1.2):  far-extended support region, i.e.\ 
$\lim_{\phi \to \pm \infty} f(\phi) \to 0$,
together with non-zero carrier envelope phase, asymmetric pulse shape,
frequency chirping, polarization gating etc. A specific class is 1.1) with
flat-top section, e.g.\ a box envelope $\sqcap$
(cf.\ \cite{King:2020hsk} for a recent explication)
belonging to $C^0$, or $\cos^2 \otimes \, \sqcap$ belonging to $C^2$ or the
construction in \cite{Otto:2014ssa} belonging to $C^\infty$. 
Examples for item 1.2) are Gauss, super-Gauss (employed in \cite{Otto:2018jbs},
for instance), symmetrized Fermi function \cite{Titov:2015pre,Titov:2014usa}, 
$1/\cosh$ etc.
The monochromatic beam, item 2), corresponds formally to an infinitely long flat-top
``pulse", abbreviated hereafter by IPA as acronym of infinite pulse approximation.
It may be considered as special case of 1.1) with $\phi_{pulse \, length \to \infty}$.
FPA stands henceforth for the finite pulse-length plane-wave approximation. 
 
To be specific, we employ here 1.2) with $f(\phi) = 1 / \cosh(\phi/\pi N)$,
where $N$ characterizes the number of oscillations in that pulse interval,
where $f(\phi)$ is significantly larger than zero
(see \cite{Kampfer:2020cbx,Titov:2015pre,Titov:2019kdk} for the formalism),
and IPA from 2). % and the constant cross field (ccf) 3). 
The laser model of class 1.1) is employed in subsection \ref{subthreshold}.

\section{Compton: differential invariant cross section %{\boldmath $d \sigma / du$}
}\label{Compton}

Let us consider the above pulse envelope function 
$f(\phi) = 1 / \cosh(\phi/(\pi N))$
to elucidate the impact of a finite pulse duration and contrast it later on
with the monochromatic laser beam model and some approximations thereof.
Differential spectra $d \sigma / du$ as a function of $u$ are exhibited in
Fig.~\ref{dsigmadu(u)} in the region $u \le 3$ for several values of $\xi \le 1$
for the FPA (dashed curves) and IPA (solid curves) models recalled below.
This complements figures 1 -- 3 in \cite{Kampfer:2020cbx}.
One observes that the harmonic structures (which would become more severe
for linear polarization, see \cite{Kampfer:2020cbx}) fade away at
larger values of $u$ and $\xi$. Therefore, we are going to analyze that region
in parameter space. There, IPA results represent reasonably well
the trends of the more involved FPA calculations. 

\subsection{Numerical results}

\begin{figure}[tb!]
\includegraphics[width=0.99\columnwidth]{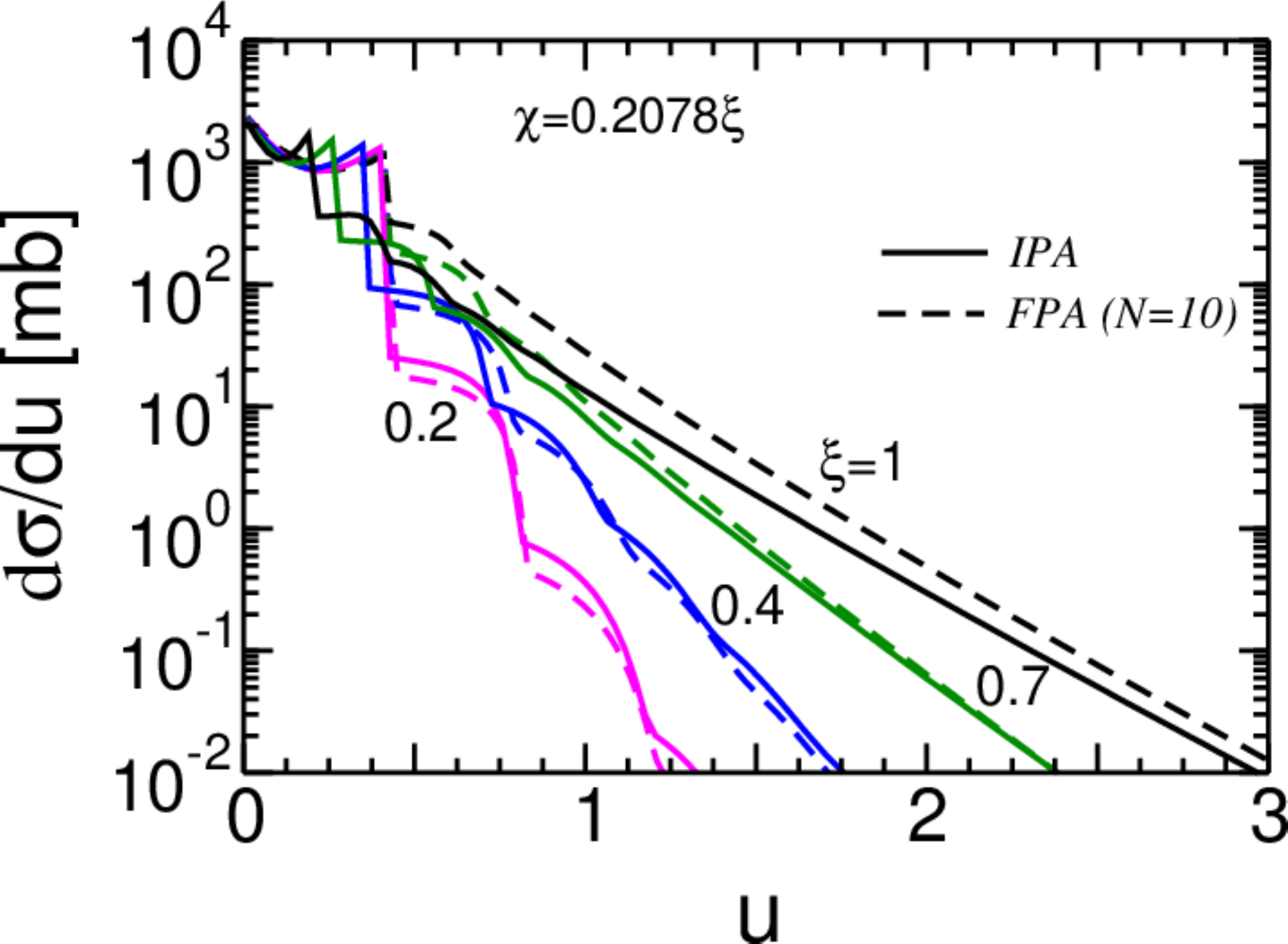}
\caption{Invariant differential cross section
$d \sigma (u, \xi, k \cdot p/m^2) / du$ 
as a function of $u$ for several values of $\xi$,
$\xi = 0.2$ (magenta), 0.4 (blue), 0.7 (green) and 1 (black).
The dashed curves (label FPA)
are for pulses according to Eq.~(\ref{laser}) with $f(\phi) = 1 / \cosh(\phi/(\pi N))$
for $N = 10$.
The IPA results for a monochromatic laser beam are depicted by solid curves
(with the limit $d \sigma_{IPA} / du\vert_{u \to 0} = 2 \pi \alpha^2 / k \cdot p$
independent of $\xi$).
The pronounced harmonic structures around the Klein-Nishina (KN) edge
$u_{KN}\approx 0.416$ are irrelevant for the subsequent discussion and, 
therefore, do not need a detailed representation.
For $\chi = \xi k \cdot p /m^2 = 0.2078 \, \xi$.
%For LUXE parameters (head-on collisions).
\label{dsigmadu(u)}}
\end{figure}

\begin{figure}[t!]
\includegraphics[width=0.99\columnwidth]{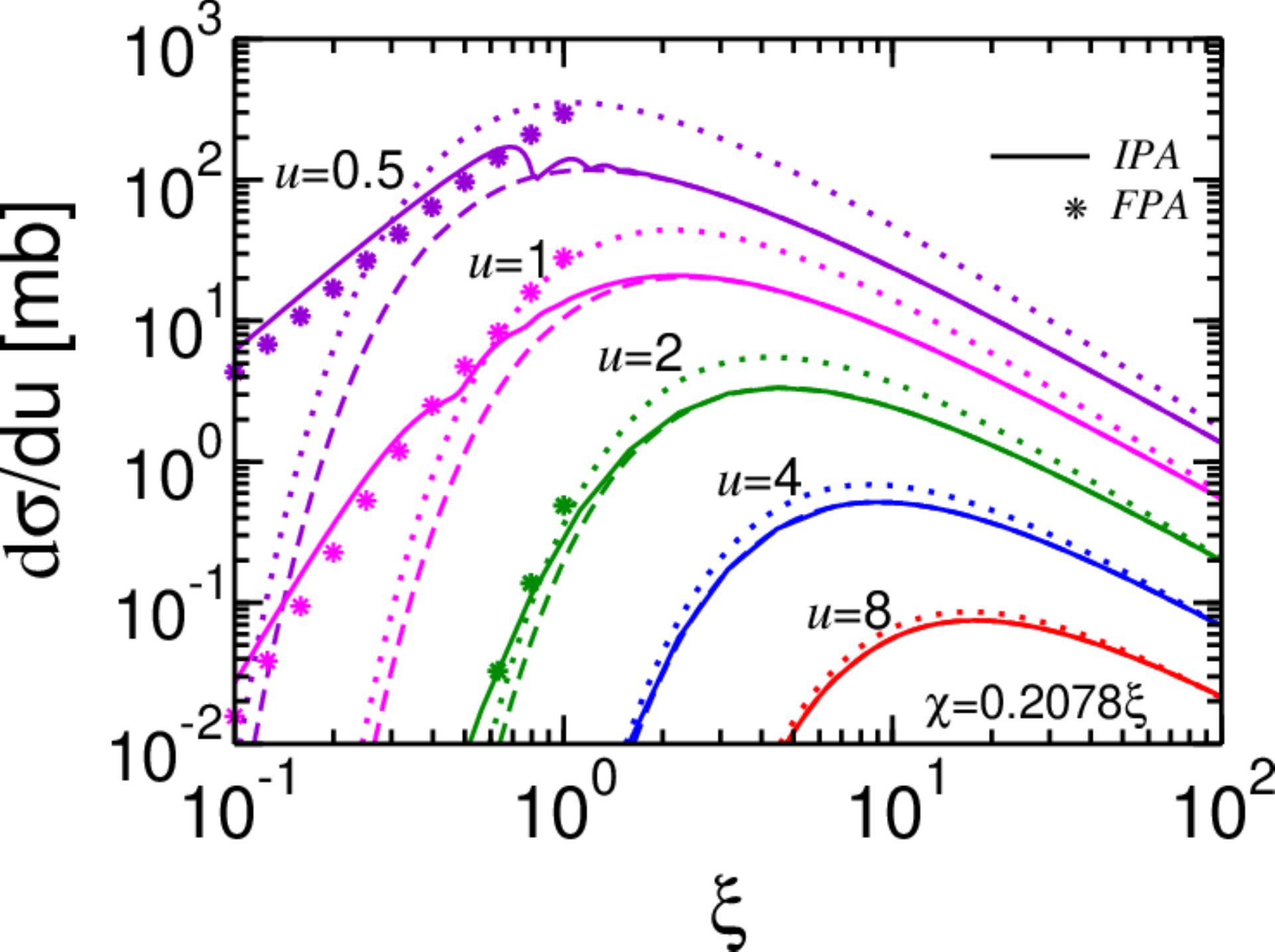}
\caption{Invariant differential cross section
$d \sigma (u = const, \xi, k \cdot p/m^2) / du$ 
as a function of  $\xi$ for several values of $u$,
$u = 0.5,$ 1, 2, 4 and 8; for $k \cdot p /m^2 = 0.2078$.
The asterisks depict results of short laser pulses with
envelope function $f(\phi) = 1 / \cosh(\phi/(\pi N))$ for
$N = 10$, $d \sigma_{FPA} / du$,  Eq.~(\ref{sigma1}).
The solid curves are for the monochromatic laser beam model
$d \sigma_{IPA} / du$,  Eq.~(\ref{IPA1}), 
while dashed (dotted) curves are based
on $d \sigma_{large-\xi} / du$,  Eq.~(\ref{ccf1})
($d \sigma_{large-\xi, \,  large-u}/du$, Eq.~(\ref{ccf2})).
%For LUXE parameters (head-on collisions).
\label{dsigma/du}}
\end{figure}

In particular, we consider now $d \sigma (u, \xi, k \cdot p/m^2) / du$ 
as a function of  $\xi$ for several constant values of $u$,
$u = 0.5,$ 1, 2, 4 and 8, for $k \cdot p /m^2 = 0.2078$ -- a value
which is motivated by the LUXE opportunities  
\cite{Abramowicz:2019gvx,Abramowicz:2021zja}. 
The solid, dotted and dashed curves in Fig.~\ref{dsigma/du}
are based on easily accessible models \cite{Ritus}:\\
$\bullet$ monochromatic model (IPA) (cf.\ equations (15, 16) in \cite{Kampfer:2020cbx}):
\begin{eqnarray} \label{IPA1}
\frac{d \sigma_{IPA}}{du} &=& \frac{ 2 \pi \alpha^2}{k \cdot p}
\frac{1}{(1+u)^2}
\sum_{n = 1}^\infty \, \Theta(u_n - u) \, F_n(z_n), \\
F_n &=&
 - \frac{2}{\xi^2} J_n^2(z_n)  \label{IPA2} \\
%\left(1 + \frac{u^2}{2(1+u)} \right) 
& + & A \left(
J_{n+1}^2(z_n) + J_{n-1}^2(z_n) - 2 J_n^2(z_n) \right) \nonumber
\end{eqnarray}
with $A = 1 + \frac{u^2}{2(1+u)}$, 
$z_n(u, u_n) = 
\frac{2 n \xi}{\sqrt{1 + \xi^2}} \sqrt{\frac{u}{u_n} (1 - \frac{u}{u_n})}$,
and $u_n = \frac{2 n k \cdot p}{m^2 (1 + \xi^2)}$
($J_n$'s denote Bessel functions of first kind);\\
$\bullet$ IPA--large-$\xi$ approximation 
(cf.\ equation (21) in \cite{Kampfer:2020cbx}):
\begin{eqnarray}  \label{ccf1} 
\frac{d\sigma_{large-\xi}}{du}
&=& - \frac{4 \pi \alpha^2}{m^2 \xi} \frac{1}{\chi} \frac{1}{(1 + u)^2} 
{\cal F}_C \\
{\cal F}_C & = &
\int_{z}^{\infty} dy \, \Phi(y) + \frac{2}{z} 
\left[1 + \frac{u^2 }{2 (1 + u)} \right]
\Phi'(z)  \label{F_C}
\end{eqnarray}
where 
$\Phi(z)$ and $\Phi(z)'$ stand
for the Airy function and its derivative
with arguments $z = (u / \chi)^{2/3}$ where $\chi = \xi k \cdot p /m^2$;\\
$\bullet$ large-$\xi$-large-$u$ approximation
(cf.\ equation (22) in \cite{Kampfer:2020cbx}):
\begin{equation} \label{ccf2}
\frac{d\sigma_{large-\xi, \, large-u}}{du} =
\frac{2 \sqrt{\pi} \alpha^2}{m^2 \xi}
\chi^{- 1/2} u^{- 3/2}
\exp \left(- \frac{2 u}{3 \chi} \right).
\end{equation}

The asterisks in Fig.~\ref{dsigma/du} depict results of 
the pulse model (FPA) described in \cite{Kampfer:2020cbx,Titov:2015pre}
(cf.\ equations (5 - 14) in \cite{Kampfer:2020cbx}):
\begin{eqnarray} \label{sigma1}
\frac{d \sigma_{FPA}}{d u} &=& \frac{\alpha^2}{k \cdot p}
\frac{1}{(1+u)^2} \int \limits_0^\infty d \ell \,
\Theta \left(\frac{u m^2}{2 k \cdot p} - \ell \right) w (\ell) , \\ 
w(\ell, z) &=&  \int_0^{2 \pi} d \phi_{e'} \left[
- \frac{2}{\xi^2} \vert \tilde Y_\ell (z_\ell) \vert^2 \label{sigma2} \right.
\end{eqnarray}
\vspace*{-6mm}
\[ +   A \left.
\left( \vert Y_{\ell - 1}(z_\ell)\vert^2 + \vert Y_{\ell +1} (z_\ell)\vert^2
- 2 \mbox{Re} \tilde Y_\ell (z_\ell) X_\ell^* (z_\ell)) \right) \right]  \] %\nonumber
with $A$ as above, 
$z_\ell (u, \ell) ={2\ell \xi}\sqrt{\frac{u}{u_\ell}(1-\frac{u}{u_\ell})}$
and $u_\ell= \frac{2 \ell k \cdot p}{m^2}$.
The basic functions $Y_\ell, \tilde Y_\ell, X_\ell$ and their $\phi_{e'}$
dependence are spelled out 
in \cite{Kampfer:2020cbx,Titov:2015pre};
they depend crucially on the pulse envelope function $f(\phi)$.
(Due to numerical accuracy reasons in integrating highly oscillating functions, 
these evaluations are constrained presently to not too large values of $\xi$.)
The common basis of these models is sketched in Appendix \ref{App_A}. 

We consider in Fig.~\ref{dsigma/du} only $u > u_{KN} = 2 k \cdot p /m^2$, 
since at the Klein-Nishina (KN) edge the
harmonic structures become severe, as seen in Fig.~\ref{dsigmadu(u)}.
The striking feature seen in 
Fig.~\ref{dsigma/du} is the pronounced 
$\cap$
%\rotatebox{180}{U}
%\raisebox{\depth}{\rotatebox{180}{U}} 
shape, which we coin ``rise and fall" of laser intensity effects. 
At small $\xi$, the realistic FPA ($N = 10$) results (asterisks) and the IPA model (solid curves)
follow the same trends, consistent with Fig.~\ref{dsigmadu(u)}. 
The large-$\xi$ and large-$\xi$-large-$u$ approximations (dashed and dotted curves) are not
supposed to apply in that region. However, they become useful representatives at large $\xi$.  \\

\subsection{The rise}

Some guidance of the rising parts of the FPA results in Fig.~\ref{dsigma/du}
can be gained by the monochromatic model. %l Eqs.~(\ref{IPA1}, \ref{IPA2})
Casting Eq.~(\ref{IPA2}) in the form
\begin{eqnarray} \label{F_n_exp}
F_n &=&
 - \frac{2 (1 - \Xi^2)}{\Xi^2} J_n^2(\Xi x_n) \\
& + & A \left(
J_{n+1}^2(\Xi x_n) + J_{n-1}^2(\Xi x_n) - 2 J_n^2(\Xi x_n) \right) \nonumber
\end{eqnarray}
with $\Xi^2 = \xi^2 / (1 + \xi^2)$ and 
$x_n (u) = 2 n \sqrt{\frac{u}{u_n} (1 - \frac{u}{u_n})}$
and expanding in powers of $\Xi$ yields for the first terms
\begin{widetext}
\begin{eqnarray} \label{F1}
F_1 &=& \frac12 (2 A - x_1^2) - \frac{x_1^2}{8} (8A -x_1^2 - 4) \Xi^2 
         + \frac{x_1^4}{384} (90 A - 5 x_1^2 - 48) \Xi^4 + {\cal O} (\Xi^6) , \\
F_2 &=&  \hspace*{2.8cm} \frac{x_2^2}{32} (8 A - x_2^2) \Xi^2 - \quad
\frac{x_2^4}{192} (18 A - x_2^2 - 6) \Xi^4 + {\cal O} (\Xi^6) , \label{F2}\\
F_3 &=& \hspace*{6.3cm} \frac{x_3^4}{1152} (18 A - x_3^2) \Xi^4
+ {\cal O} (\Xi^6) . \label{F3}
\end{eqnarray}
\end{widetext}
Due to the Heavyside $\Theta$ function in Eq.~(\ref{IPA1}), the
leading-order power in $\Xi$ depends on the value of $u$, e.g.\
for $u < u_1$, the series starts with ${\cal O} (\Xi^0)$ and the 
coefficients of higher orders accordingly sum up. Higher values of $u$
facilitate higher orders of the leading terms, i.e.\ the rise becomes steeper
since the respective leading-order term is $\propto \Xi^{2 \lfloor u/u_1 \rfloor}$.
This statement is based on the structures in Eqs.~(\ref{F1} - \ref{F3}),
suggesting $F_n = \sum_{i = n - 1} F_{n}^{(2i)} \, \Xi^{2 i}$ 
for the first terms $F_{n}^{(2i)}$, and 
$\sum_{n=1}^\infty \Theta (u_n - u) \, F_n = \sum_{n_{min}}^\infty F_n$
with $n_{min} = 1 + \lfloor u/u_1 \rfloor$. 
($\lfloor \cdot \rfloor$ is the floor operation.)

The series expansion of (\ref{F_n_exp}) in powers of $\Xi$ ignores the sub-leading $\Xi$ dependence
in $x_n$ via $u_n = 2 n \frac{k \cdot p}{m^2} (1 - \Xi^2)$ but is suitable for $k \cdot p = const$. 

Analogously considerations apply to the pulse model,
cf.\ section III.C in \cite{Titov:2015pre}.
An essential role is played by the Fourier transform of the 
pulse envelope in the limit $\xi \ll 1$. It bridges to the IPA
for long pulses. 

\subsection{The fall}

The maximum of the curves exhibited in Fig.~\ref{dsigma/du}
at $u = {\cal O}(1)$ is attained at $\xi = {\cal O}(1)$,
but moves towards larger values of $\xi$ with increasing values of $u$. 
Remarkably, 
the often discredited large-$\xi$ and large-$\xi$--large-$u$
approximations (dotted and dashed curves) 
%and their equivalent ccf expressions 
deliver results
in fair agreement with the IPA results (solid curves) for $u > 1$.
That is, when being interested in the high-energy photon tails, the simple
large-$\xi$--large-$u$ formula Eq.~(\ref{ccf2})  \cite{Ritus}
represents a fairly accurate description supposed $\xi$ is sufficiently large.
Obviously, at $\xi < 1$ and $u < 2$, such an approximation fails quantitatively. 
(In particular, at $u < 1$ the harmonic structures in IPA become severe.)
Nevertheless, some estimate of the maximum position is provided by
$\xi_{max} \approx \frac49 u m^2 / k \cdot p$ yielding
$d \sigma / du (\xi_{max}) \approx \frac{2 \sqrt{\pi} \alpha^2}{m^2}
\frac{k \cdot p}{m^2} \left( \frac32 \right)^3 e^{- 3/2} \, u^{-3}$.
The asymptotic fall is governed by
$\frac{d \sigma}{du} \propto \frac{1}{m^2} \frac{m^2}{k \cdot p}
\xi^{-3/2} u^{- 3/2} \exp \left( - \frac{2 u m^2}{3 \xi k \cdot p}\right)$
from Eq.~(\ref{ccf2}).

\begin{figure}[b!]
\includegraphics[width=0.99\columnwidth]{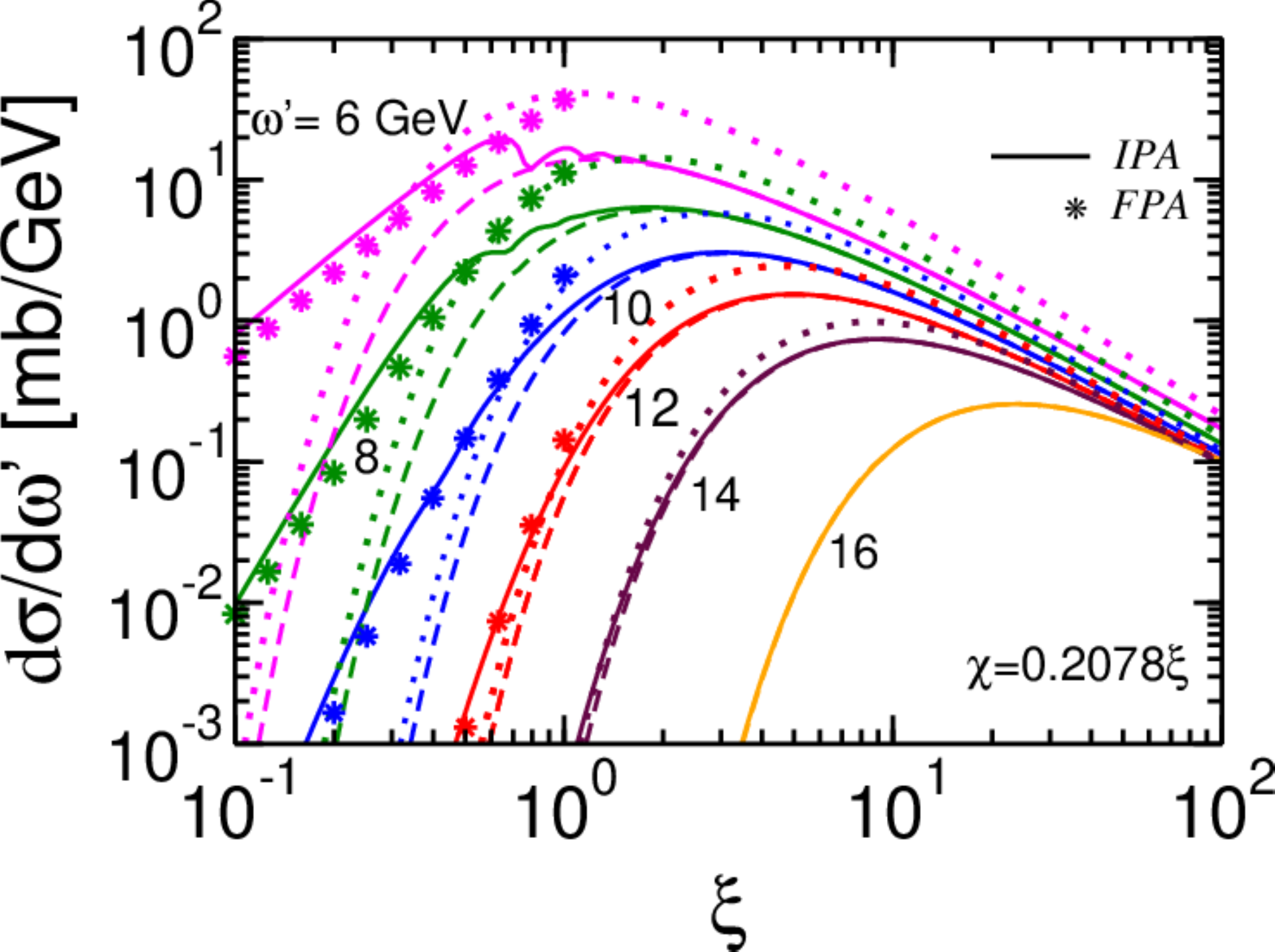}
\caption{The same as in Fig.~\ref{dsigma/du} but for $d \sigma /d \omega'$ as a function
of laser intensity parameter $\xi$ for several constant values of $\omega'$.
%For LUXE parameters (head-on collisions).
\label{dsigma/domegap}}
\end{figure}

We emphasize the same pattern of rise and fall of $d \sigma /d \omega' \vert_{\omega' = const}$,
see Fig.~\ref{dsigma/domegap}, again for sufficiently large values of $\omega'$.
The spectrally resolved differential cross section $d \sigma /d \omega' \vert_{\omega' = const}$
is directly accessible in experiments. In the strong-field asymptotic region it displays a funneling behavior,
i.e.\ the curves are squeezed into a narrow corridor already in the non-asymptotic $\xi$ region,
in contrast to $d \sigma /d u\vert_{u= const}$ in Fig.~\ref{dsigma/du}.

\section{Discussion} \label{discussion}

\subsection{Cross section vs. probability} \label{prob_cc}

The cross section $\sigma$ and Ritus probability rate $W$ are related as 
$W =  \frac{m^4}{4 \pi \alpha} \frac{\xi \, \chi}{q_0} \sigma $
with $q_0$ denoting the energy component of the quasi-momentum of the $in$-electron.
This relation holds true for circular polarization
and applies to respective differential quantities too.
The different normalization modifies in particular the $\xi$ dependence: The above
emphasized ``rise and fall" of $d \sigma /du$ corresponds to a
monotonously rising probability
%$d \mathbb{P} /du$ 
$d W / du$ as a function of $\xi$. Having in mind Ritus' remark
``the cross-sectional concept becomes meaningless" since, at $\xi \to \infty$
we have
$\sigma \to 0$ while $W$ remains finite \cite{Ritus}, we turn in this
sub-section to the probability. In doing so we remind the reader of the subtle Ritus notation
$W(\chi) = \frac{1}{\pi} \int_0^{\pi} d \psi P(\chi \sin \psi)$ in distinguishing the probabilities
$W$ and $P$. 

We also stress the varying behavior of total 
(cf.\ appendix A in \cite{Titov:2010ps})
and differential probabilities. 
Equation (\ref{ccf2}) emerges as a certain limit of the
constant cross field probability \cite{Ritus}
$\frac{d P (\chi, u)}{du} = - V %\frac{\alpha m^2}{ \pi q_0} 
(1 + u)^{-2} \, {\cal F}_C (z(\chi, u), u)$
where the prefactor reads 
$V = \alpha m^2 / \pi q_0$ and
${\cal F}_C$ is defined in Eq.~(\ref{F_C}). 
With respect to the argument  $z = (u/\chi)^{2/3}$
of the Airy functions in Eq.~(\ref{F_C}),
the $\chi$ - $u$ plane can be divided into
the regions I (where $u > \chi$) and II (where $u < \chi$). Furthermore,
the combination $1 + u$ suggests a splitting into $u < 1$ and $u > 1$
sub-domains. Picking up the leading-order terms in the related series expansions 
one arrives at 
\begin{eqnarray}
\frac{d P}{du}\vert_{u \gg \chi} 
&=& \frac{V}{2 \sqrt{\pi}} \chi^{1/2}
\exp \left( - \frac{2u}{3 \chi} \right) \nonumber \\
& \times &
\left\{
\begin{array}{l}
u^{- 1/2} \quad \mbox{for} \quad u \ll 1, \quad (I^{<})\\
u^{- 3/2} \quad \mbox{for} \quad u \gg 1,  \quad (I^{>})\\
\end{array} 
\right. \label{region_I} \\
\frac{d P}{du}\vert_{u \ll \chi} 
&=& - V \Phi'(0) \chi^{2/3}  \nonumber \\
& \times &
\left\{
\begin{array}{l}
2 u^{- 2/3} \quad \mbox{for} \quad u \ll 1,   \quad (II^{<})\\
u^{- 5/3}  \,\,\,\quad \mbox{for} \quad u \gg 1.   \quad (II^{>})\\
\end{array} 
\right. \label{region_II}
\end{eqnarray}
It is the soft $u$-part $II^{<}$ of the differential probability in Eq.~(\ref{region_II})
which essentially determines the total probability upon $u$-integration,\footnote{
The value $\chi = 1$ is special since the small-$u$ region of II and the large-$u$ region of I must be joint directly, $II^{<} \otimes I^{>}$,
while for $\chi < 1$ the small-$u$ region II
and the small-$u$ region of I must be joint followed by the large-$u$ region of I,
$II^{<} \otimes I^{<} \otimes I^{>}$.
In the opposite case of $\chi > 1$, the small-$u$ region of II must be joint 
with the large-$u$ region of II followed by the large-$u$ region of I,
$II^{<} \otimes II^{>} \otimes I^{>}$. This
distinction becomes also evident when inspecting the differently shaped sections
of the continuous curves $d P /du$ as a function of $u$ for several values of $\chi$. }
i.e.\ the celebrated result $P \propto \alpha \chi^{2/3}$ in Ritus notation \cite{Ritus} 
(stated as 
$\lim_{\chi, \xi \to \infty} \mathbb{P} 
\propto \alpha \chi^{2/3} m^2 / k \cdot p$ in \cite{Ilderton:2019kqp}), 
while the hard $u$-part $I^{>}$ in Eq.~(\ref{region_I}) 
is at the origin of Eq.~(\ref{ccf2}) when converting to cross section. In other words, one has to distinguish
the $\xi$ (or $\chi$) dependence either of integrated or differential observables.
In relation to the LUXE plans \cite{Abramowicz:2019gvx}
we mention the photon detector developments \cite{Fleck:2020opg}
which should enable in fact the access to the differential spectra.

\subsection{Secondary processes: Breit-Wheeler}\label{BW}

Instead transferring these considerations to the Breit-Wheeler (BW)
process {\it per se} (cf.\ 
\cite{Kaminski:2006xlq,Heinzl:2010vg,Krajewska:2012eb,Krajewska:2014ssa,Jansen:2016gvt,Jansen:2016crq,Golub:2020kkc,Titov:2020taw} 
and further citations below), 
we estimate now the BW pair production seeded by the hard photons
from the above Compton (C) process. The following folding model 
C$\otimes$BW is a pure two-step ansatz on the probabilistic level
which mimics in a simple manner
some part of the trident process\footnote{For recent work on the
formalism, see
\cite{Torgrimsson:2020wlz,Dinu:2019wdw,Dinu:2019pau,King:2018ibi,Mackenroth:2018smh,Acosta:2019bvh}
which advance earlier investigations \cite{Baier:1972vc,Ritus:1972nf}.} 
$e_L^-  \to {e_L^-}' + {e_L^-}'' + e_L^+$
by ignoring (i) the possible off-shell effects and non-transverse components 
of the intermediate photon (that would be the one-step contribution) and (ii) the
anti-symmetrization of the two electrons in the final state
(that would be the exchange contribution). 
Such an ansatz is similar to the one in \cite{Hartin:2018sha}, where however
bremsstrahlung$\otimes$BW has been analyzed. An analog approach has been
elaborated in \cite{Titov:2009cr} for di-muon production.

Specifically, we consider the two-step cascade process where
a GeV-Compton photon with energy $\omega'$ is produced
%$e-L$ interactions ($L$ means laser pulse) 
in the first step,
and, in the next step, that GeV-photon interacts with the same laser field 
producing a BW-$\ee$ pair. 
We estimate the number of pairs produced in one pulse by
\begin{equation} \label{pair_number}
N^{\ee} = \int\limits_0^\infty d \omega' \,
\int\limits_0^{T_C} dt \, \frac{d \Gamma_C (\omega', t)}{d \omega'}
\int\limits_t^{T_{BW}} dt' \, \Gamma_{BW} (\omega', t') ,
\end{equation}
where $d \Gamma_C (\omega', t) / d \omega'$ is the rate 
of photons per frequency interval $d \omega'$ 
(which corresponds to $d P /du$  in \ref{prob_cc})
emerging from Compton at time $t \in [0, T_C]$, and  
$\Gamma_{BW} (\omega', t')$ is the rate of Breit-Wheeler pairs generated
by a probe photon of frequency $\omega'$ at lab.\ frame time-distance 
$t' \in [t, T_{BW}]$. 
The underlying picture is that of an electron traversing
a laser pulse in head-on geometry near to light cone.
The passage time of an undisturbed electron
would be $N T_0/2$, $T_0 = 2 \pi / \omega$. 
Neglecting spatio-temporal variations within the pulse, 
the final formula becomes 
$N^{\ee} = F_t \int_0^{E_e} d \omega' \, N^e_0
\frac{d \Gamma_C (\omega')}{d \omega'} \, \Gamma_{BW} (\omega')$
upon the restriction $\omega' < E_e$ and 
$F_t = T_C (T_{BW} - T_C / 2)$.
A crucial issue is the choice of the formation time(s) \cite{Baier:1972vc,Ritus:1972nf}.
When gluing C$\otimes$BW on the amplitude level, such a time
appears linearly in the pair rate \cite{Hu:2010ye}\footnote{
In laser pulses such an additional parameter is not needed 
\cite{Ilderton:2010wr}.}
or quadratically in the net probability via overlap light-cone times
(cf.\ \cite{King:2018ibi,Mackenroth:2018smh} for instance).
The time-ordered double integral over the C and BW probabilities, yielding 
the cascade approximation (cf.\ equation (43) in \cite{Mackenroth:2018smh}), 
is analog to our above formula by folding two rates,  
which facilitates $F_t \propto T_0^2$, if $T_{C, \, BW} \propto T_0$.
We attach to the Compton rate the number $N^e_0 = 6 \times 10^9$ of
primary electrons per bunch.

\begin{figure}[!tb]
\includegraphics[width=0.99\columnwidth]{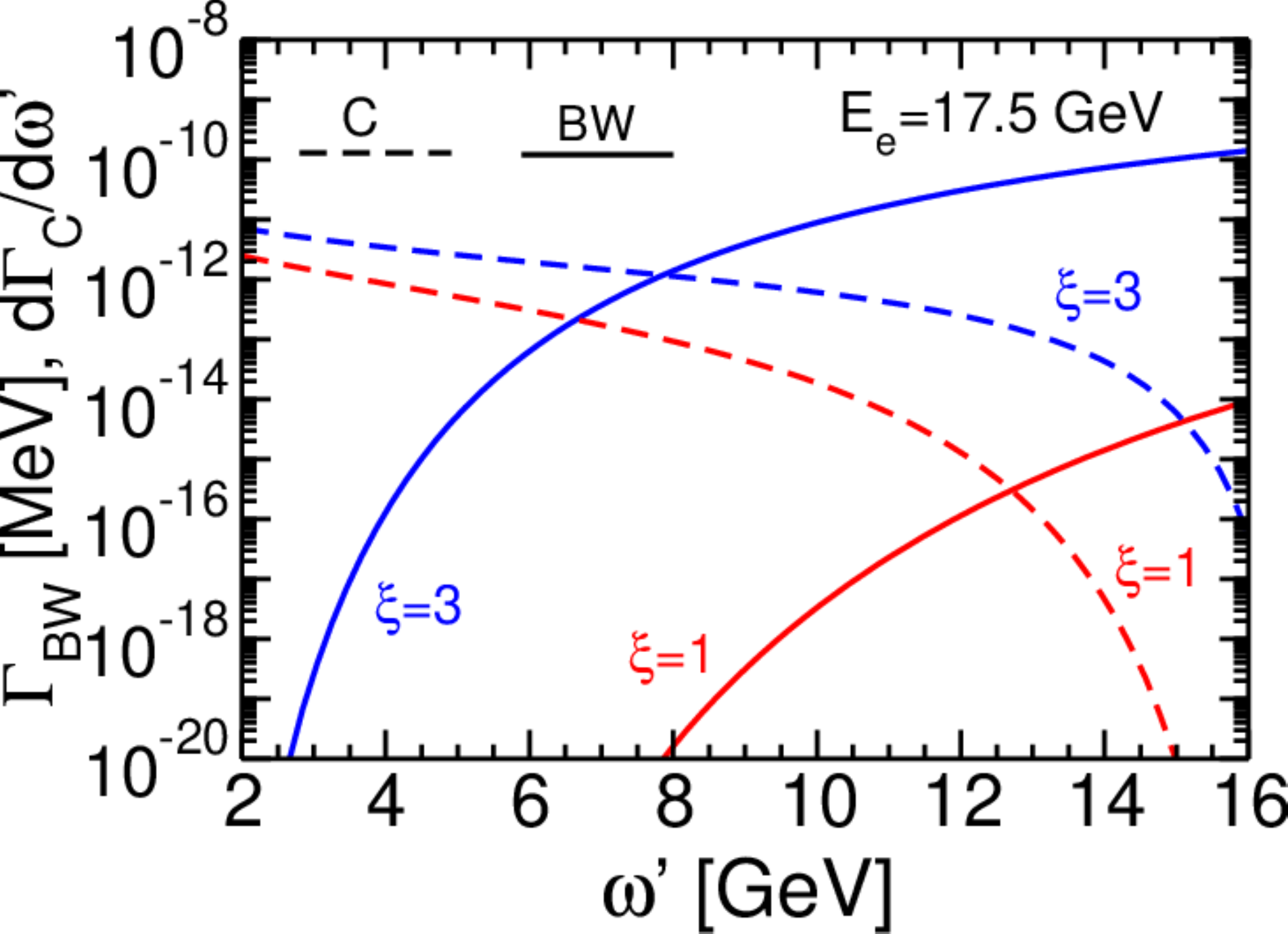}
\caption{The BW rate  $\Gamma_{BW}$ from Eqs.~(\ref{G_BW}, \ref{BW_rate}) (solid curves) 
and the  dimensionless differential C rate $d \Gamma_C / d \omega'$ 
(dashed curves) from
%Eqs.~(\ref{rate_C_IPA}, \ref{rate_C_IPA_cont})
Eqs.~(\ref{F_C}, \ref{R1})
as a function of $\omega'$ for  $\xi = 1$ (red) and 3 (blue).
\label{CxBW}
}
\end{figure}

In the numerical evaluation we employ the following convenient approximations:
\begin{eqnarray} \label{R1}
\frac{d \Gamma_C}{d \omega'} &=&
-  \frac{\alpha m^2}{\pi \,E_e^2}   
{\cal F}_C(z, u),\\
\Gamma_{BW} (\omega') &=& \label{G_BW}
\frac{\alpha m^2}{\omega'} {\cal F}_{BW} , \\
{\cal F}_{BW} &=& \sqrt{\frac{3^3}{2^9}}\kappa 
\exp \left[-\frac{8}{3\kappa} \left(1-\frac{1}{15\xi^2}
\right) \right] ,
\label{R3}
\end{eqnarray}
where $u=\kappa /(\chi - \kappa)$, $\chi=(k\cdot p)\xi/m^2$,
$\kappa=(k\cdot k')\xi/m^2$, 
%$\Phi(z)$ and $\Phi(z)'$ stand for the Airy function and its derivative
and ${\cal F}_C (z,u)$ with $z = (u / \chi)^{2/3}$
is defined in Eq.~(\ref{F_C}).
Note the rise of $\Gamma_{BW}$ and the fall of 
$d \Gamma_C / d \omega'$ as a function of $\omega'$
at fixed values of $\xi$ and $k \cdot p$ and laser frequency $\omega$,
see Fig.~\ref{CxBW}. Increasing values of $\xi$ lift $d \Gamma_C / d \omega'$
somewhat and make it flatter in the intermediate-$\omega'$ region, 
thus sharpening the drop at $\omega' \to E_e$.
Remarkable is the strong impact of increasing $\xi$ on the BW rate.

One may replace Eq.~(\ref{F_C}) in (\ref{R1}) by the sum over harmonics expressed
by Bessel functions \cite{Ritus} to get an improvement of accuracy in the small-$\omega'$
region:\footnote{Strictly speaking, imposing a finite duration in the monochromatic
laser beam model  2) turns it into a flat-top laser pulse model of class 1.1),
exemplified in \cite{King:2020hsk} and applied to the nonlinear Compton process.}
\begin{eqnarray}
\frac{d \Gamma_C^{IPA}}{d \omega'} &=& \label{rate_C_IPA}
\frac{d \sigma_C^{IPA}}{d \omega'}
\frac{\xi^2 m^2}{4 \pi \alpha} \frac{k \cdot p}{q_0} ,\\
\frac{d \sigma_C^{IPA}}{d \omega'} &=& \frac{\pi r_e^2 m^2}{\xi^2 k \cdot p}
%{\cal I} \\
%{\cal I} &=&  
\sum_{n=1}^\infty \frac{\Theta (u_n - u)}{P_n} 
\left[ - 4 J_n^2 \right. \label{rate_C_IPA_cont} \\
&+& \left. \xi^2 (2 + \frac{u^2}{1+u})
\left(J_{n-1}^2 + J_{n+1}^2 - 2 J_n^2 \right) \right] \nonumber
\end{eqnarray}
with $q_0 = \sqrt{E_e^2 - m^2} + \beta_p \omega$,
$\beta_p = \xi^2 \frac{m^2}{2 q \cdot k}$,
$q \cdot k = k \cdot p$, and arguments 
$z = \frac{2 n \xi}{\sqrt{1+\xi^2}} \frac{1}{u_n} \sqrt{u (u_n-u)}$
of the Bessel functions $J_n$ as well as
$u_n = 2 n \frac{k \cdot p}{m^2} \frac{1}{1 + \xi^2}$,
$u = \frac{n \omega - \omega'}{\kappa_n - n \omega + \omega'}$,
$\kappa_n = n \omega - \frac12 m e^\zeta + \frac12 m (1+\xi^2) e^{- \zeta}$,
$P_n = m \vert n \frac{\omega}{m} - \sinh \zeta 
+ \frac{\xi^2}{2} e^{-\zeta} \vert$.
These equations make the dependence $u(\omega')$ explicit and relate again
the differential  cross section
$d \sigma / d \omega'$ with the differential rate $d \Gamma / d \omega'$. 
Since the BW rate is exceedingly
small at small $\omega'$ (see Fig.~\ref{CxBW}), 
improvements of the Compton rate by catching the details of harmonic structures
there are less severe for the pair number Eq.~(\ref{pair_number}).

Equation (\ref{R3}) of the BW rate, however, is inappropriate at smaller values of $\xi$ 
\cite{Hartin:2018sha}
and needs improvement. Instead of using the series expansion in Bessel
functions \cite{Ritus}, a convenient formula is
\begin{eqnarray}
{\cal F}_{BW} &=& \frac{1}{4 \pi} \label{BW_rate}
\sum_{n = n_{min}}^\infty
\int_1^{u_n} \frac{du}{u^{3/2} \sqrt{1 - u}} \\
& \times &
\exp\left\{ - 2 n (a - \tanh a \right\} 
\frac{1 + 2 \xi^2 (2u-1) \sinh^2 a}{a \tanh a} \nonumber
\end{eqnarray} 
with $n_{min} = 2 \xi (1 + \xi^2) / \chi$ and $u_n = n / n_{min}$.
This representation emerges from the large-$n$ approximation of Bessel functions,
$J_n(z) \approx \exp\left\{ - n (a- \tanh a) \right\} / \sqrt{2 n \pi \tanh a}$ and 
$\tanh a = \sqrt{1 - z^2 / n^2}$. In the large-$\xi$ limit, one may replace the
summation over $n$ by an integration to arrive, via a double saddle point
approximation, at the famous Ritus expression (\ref{R3}), 
which in turn is a complement of Eq.~(\ref{ccf2}), see \cite{Ritus}.

\begin{figure}[!t]
\includegraphics[width=0.99\columnwidth]{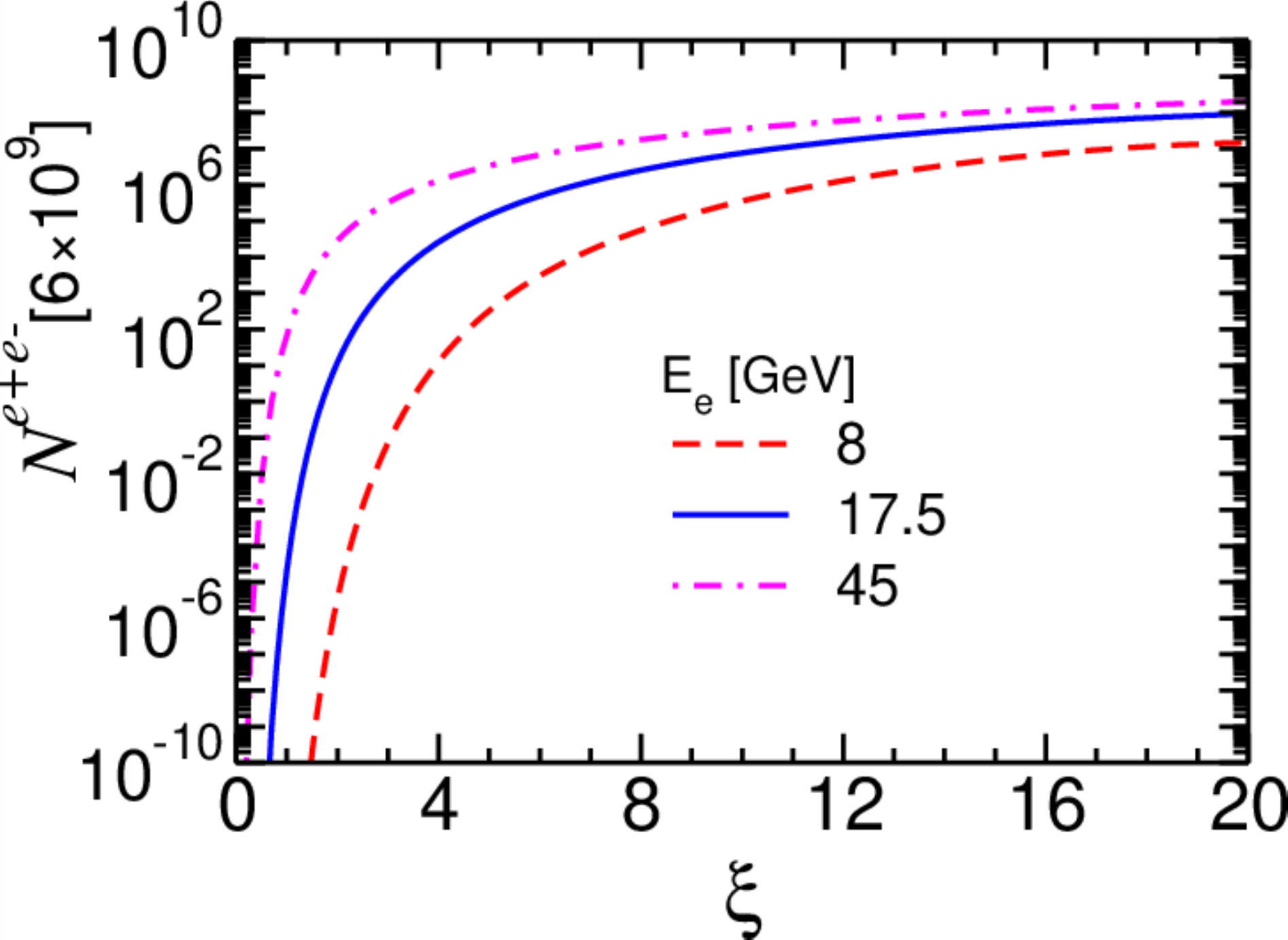}
\caption{The yield of $\ee$ pairs as a function of $\xi$ for
electron energies 45~GeV (magenta dash-dotted curve),
17.5~GeV (blue solid curve) and 8~GeV (red dashed curve)
according to the probabilistic folding model C$\otimes$BW
Eq.~(\ref{pair_number}). 
For $N_0^e = 6 \times 10^9$ electrons per bunch and per laser shot of duration $N T_0$.
The special normalization $F_t = T_0^2/2$ is chosen,
as realized by $T_{BW} = (T_0^2 + T_C^2)/(2 T_C)$. The choice 
$T_C \approx T_0$ facilitates a Compton spectrum
$d N_C/d \omega' = T_C \, d \Gamma_C / d \omega'$ which agrees, 
for $\xi = {\cal O} (1)$ and
in the region $\omega' < 10$~GeV, with a bremsstrahlung spectrum 
generated by electrons of the same energy impinging on a foil with 
$X/X_0 = 0.01$ \cite{Hartin:2018sha}. 
It is the $\xi$ dependence of the Compton spectrum
(see  Fig.~\ref{CxBW}) which makes the pair yield more rapidly rising 
with $\xi$ than the
pair yield of the bremsstrahlung$\otimes$BW model in \cite{Hartin:2018sha}.
\label{Ratio}}
\end{figure}

\begin{figure}[!h]
\includegraphics[width=0.95\columnwidth]{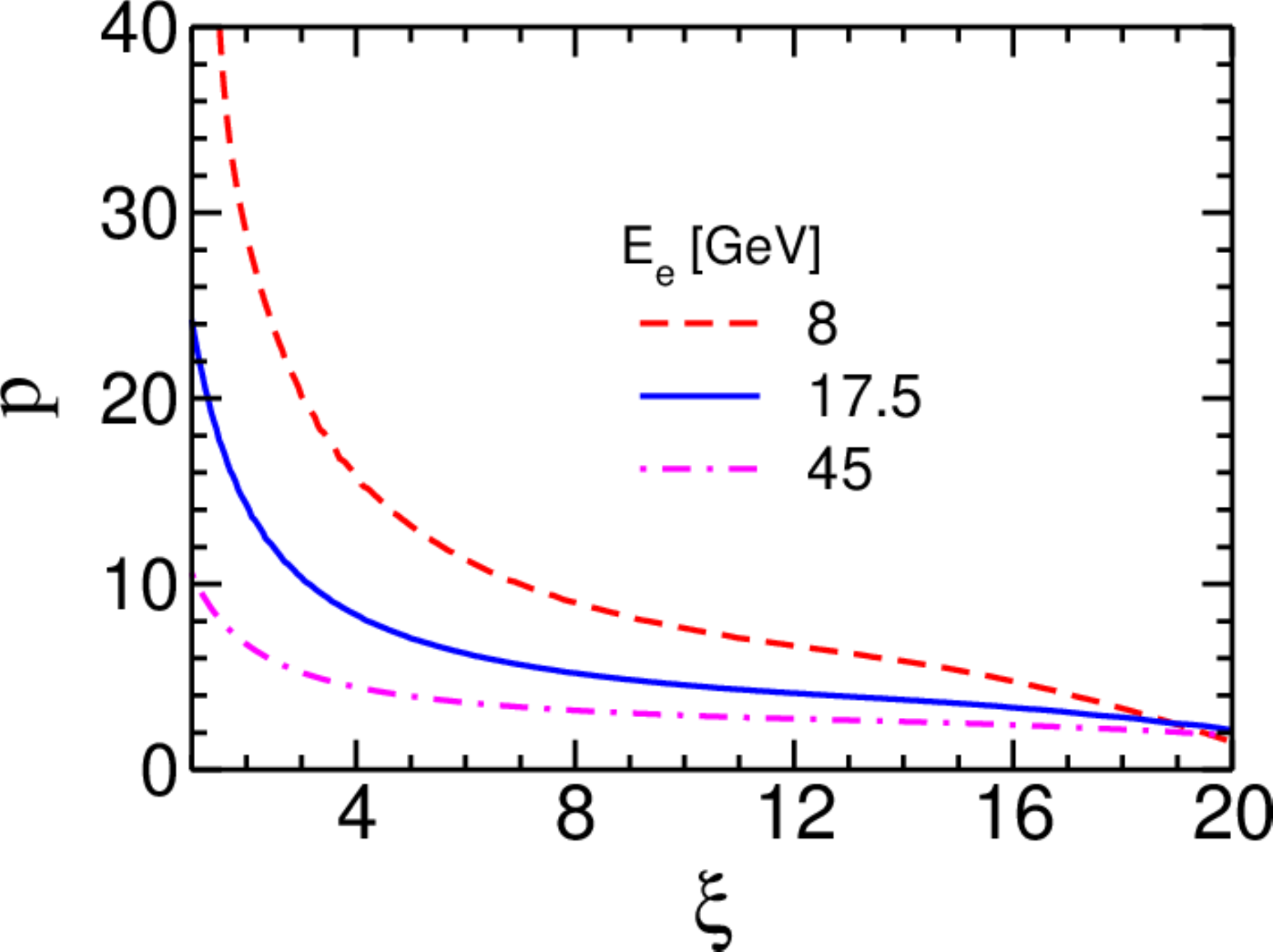}
\caption{The power $p$ as a function of $\xi$ for $E_e = 45$~GeV
(magenta dash-dotted curve), 17.5~GeV (blue solid curve) and 8~GeV
(red dashed curve). The results exhibited in Fig.~\ref{Ratio} are described by 
$N^{\ee} (\xi, E_e) = N^{\ee}_0 (E_e)  \, \xi^{p(\xi, E_e)}$.
\label{power_p}}
\end{figure}

Numerical results are exhibited in Fig.~\ref{Ratio} for $E_e = 45$~GeV,
17.5~GeV and 8~GeV. 
One observes a stark rise of $N^{e^+ e^-}$ up to $\xi \sim 4$, which 
turns for larger values of $\xi$ into a modest rise. 
To quantify that rise one can employ the ansatz 
$N^{\ee} (\xi, E_e) = N^{\ee}_0 (E_e)  \, \xi^{p(\xi, E_e)}$.
Note that, by such a quantification of the $\xi$ dependence, one gets rid of the normalization $F_t$.
For $E_e = 17.5$~GeV we find $p(\xi \approx 1) \approx 20$ dropping to 
$p(\xi \approx 20) \approx 2$, see Fig.~\ref{power_p}.
Larger values of $E_e$ reduce $p$, 
e.g.\ $p(\xi \approx 1)\vert_{E_e = 45 \, \rm{GeV}} \approx 10$ 
in agreement with \cite{Hu:2010ye}, while 
$p(\xi \approx 1)\vert_{E_e = 8 \, \rm{GeV}} > 40$. 
At $\xi \to 20$, a universal value of $p \approx 2$ seems
to emerge.
The extreme nonlinear sensitivity of the
pair number on the laser intensity parameter $\xi$ at $\xi < 10$,
and in particular at $\xi \approx 1$,
points to the request of a refined
and adequately realistic modeling beyond schematic approaches.

\subsection{Bandwidth effects in linear trident}\label{subthreshold}

\begin{figure}[!tb]
\includegraphics[width=0.99\columnwidth]{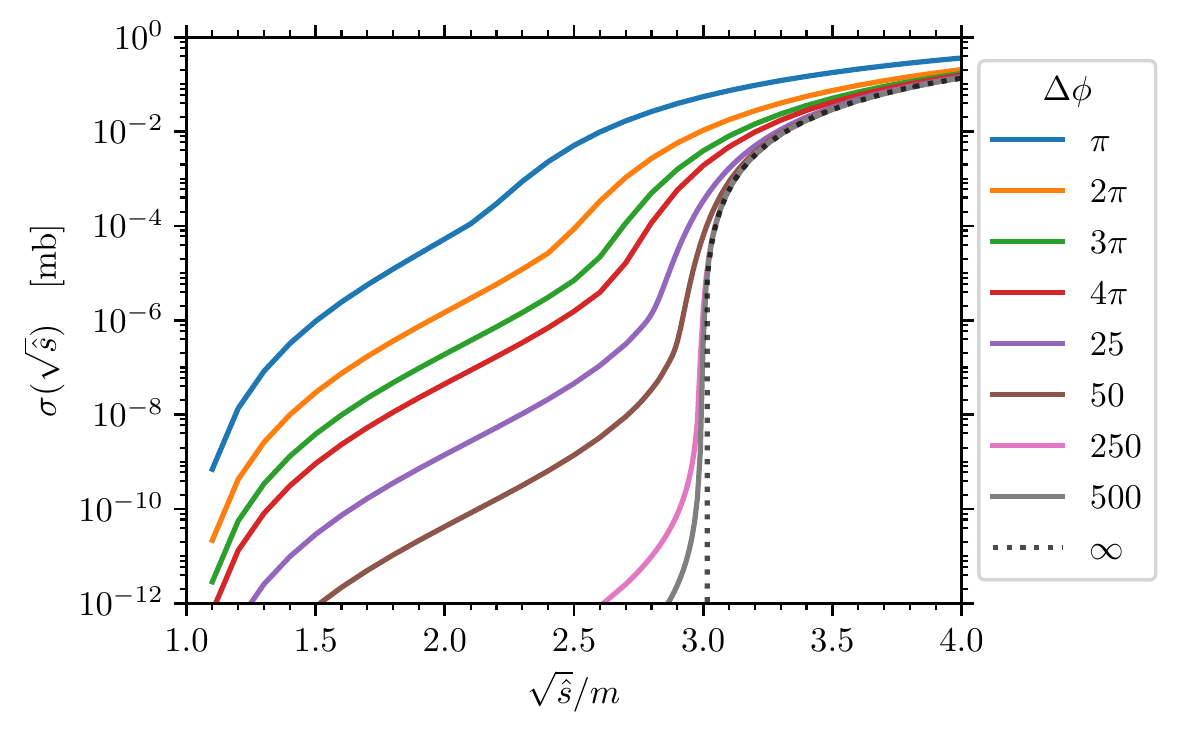}
\caption{Pulsed perturbative trident cross section $\sigma_{ppT}$ as a function
of $\sqrt{\hat s} / m$ for several pulse lengths $\Delta \phi$. In the IPA limit,
i.e.\ a monochromatic laser beam or $\Delta \phi \to \infty$, the threshold
is at $\sqrt{\hat s} = 3 m$.  Above the threshold, the dots depict a few points
(cf.\ table 1 in \cite{Haug:1975bk})
from perturbative trident without bandwidth effects.
Bandwidth effects enable the pair production in the sub-threshold region 
$\sqrt{\hat s} < 3 m$.
}
\label{fig:ppT}
\end{figure}

The threshold for linear trident,
$e^- + \gamma(1.55 \, {\mbox eV}) \to {e^-}' + {e^-}'' + e^+$,
is at $E_e = 337$~GeV, i.e.\ the LUXE kinematics is in the deep sub-threshold
regime, where severe multi-photon effects build up the nonlinearity.
However, also bandwidth effects can promote pair production
in the sub-threshold region \cite{Titov:2012rd,Nousch:2012xe},
even at $\xi \to 0$. 
The key is the cross section of linear
trident $\sigma_{ppT} (\sqrt{\hat s}, \Delta \phi)$, which depends on the invariant
energy $\sqrt{\hat s} = m \sqrt{1 + 2 k \cdot p/m^2}$ and the pulse duration 
$\Delta \phi$ for a given laser pulse. The quantity   
$\sigma_{ppT} (\sqrt{\hat s}, \Delta \phi)$ is exhibited in
Fig.~\ref{fig:ppT} as a function
of $\sqrt{\hat s}$ for several values of $\Delta \phi$. For definiteness, we employ
the laser pulse model of class 1.1) with parameterization 
$\vec A = f_{ppT} (\phi) \, \vec a_x \, \cos \phi$ and envelope function 
$f_{ppT} = \cos^2 \left( \frac{\pi \phi}{2 \Delta \phi} \right)
\sqcap (\phi, 2 \Delta \phi)$, i.e.\ the number
of laser-field oscillations within the pulse is $N = \Delta \phi / \pi$. 
In contrast to the presentation above, we deploy results in this sub-section
for linear polarization and the $\cos^2$ envelope. 

We employ the formalism in 
\cite{Acosta:2019bvh} and its numerical implementation, that is 
``pulsed perturbative QED" in the spirit of Furry picture QED in a
series expansion in powers of $\xi$. Applied to trident, the
pulsed perturbative trident (ppT) arises from the diagrams
\setlength{\unitlength}{1mm}
\thicklines
\begin{picture}(8,8)
\put(0,0.5){\line(1,0){8}}
\put(0,1.0){\line(1,0){8}}
\put(3.0,0.5){\line(0,1){5.5}}
\put(3.0,5.5){\line(1,0){5}}
\put(3.0,5.9){\line(1,0){5}}
\put(3.0,5.5){\line(2,-1){5}}
\put(3.0,4.8){\line(2,-1){5}}
\put(9,4.7){{\footnotesize $e^+$}}
\put(9,2.2){{\footnotesize $e^-$}}
\put(3,0.5){\circle*{1}}
\put(3,5.3){\circle*{1}} 
\end{picture}
\begin{picture}(8,8)
\put(4.5,2.0){$-$}
%\put(0,2.0){$-$}
\end{picture}
%$-$
\begin{picture}(8,8)
\put(0,0.5){\line(1,0){8}}
\put(0,1.0){\line(1,0){8}}
\put(3.0,0.5){\line(0,1){5.5}}
\put(3.0,5.5){\line(1,0){5}}
\put(3.0,5.9){\line(1,0){5}}
\put(3.0,5.5){\line(2,-1){5}}
\put(3.0,4.8){\line(2,-1){5}}
\put(9,4.7){{\footnotesize $e^-$}}
\put(9,2.2){{\footnotesize $e^+$}}
\put(3.1,0.5){\circle*{1}} 
\put(3.1,5.3){\circle*{1}} 
\end{picture}
%\hspace*{2mm}
(double lines: Volkov wave functions, vertical lines: photon propagator)
as leading-order term surviving $\xi \to 0$.
\setlength{\unitlength}{1pt}

The scaled number of pairs is $N_{tot} / N_0^e= 2 \pi \Gamma_{tot} / \pi$,
where the probability rate is given by 
\begin{equation}\label{Wtot}
\Gamma_{tot} = \sigma_{ppT} \frac{\omega}{2} \frac{m^2 \xi^2}{4 \pi \alpha}
\int_{-\infty}^{\infty} d \phi \, f^2_{ppT} (\phi).
\end{equation}
The chosen pulse implies $\int_{-\infty}^{\infty} d \phi \, f^2_{ppT} (\phi) =
\frac32 \Delta \phi$ and 
$N_{tot} / N_0^e = \sigma_{ppT} \frac{3 m^2 \xi^2}{16 \alpha} \Delta \phi$.
Note the $\xi^2$ dependence from the ``target density" already
entering in Eq.~(\ref{Wtot}) (cf.\ \cite{Seipt:2010ya,Titov:2012rd}
for analog relations).
This is in contrast to Fig.~\ref{Ratio},
where genuine nonlinear effects are at work
and mix with a stronger $\xi$ dependence
for C$\otimes$BW.
The $\xi^2$ dependence is characteristic for pair production by probe photons
provided by an ``external target", such as in the bremsstrahlung-laser
configuration of LUXE, cf.\ \cite{Hartin:2018sha}.

For the long laser pulses used in E-144 
\cite{Bamber:1999zt,Burke:1997ew,Bula:1996st}, 
such bandwidth effects are less severe.

\section{summary}\label{summary}

In summary, inspired by the renewed interest in the Ritus-Narozhny conjecture
and the new perspectives offered by the experimental capabilities of LUXE
and E-320, we recollect a few features of elementary QED processes
within the essentially known formalism.
In particular, we focus on the $\xi$ dependence.
For nonlinear Compton scattering,
we point out that, in the non-asymptotic region
$\chi = {\cal O}(1)$, $k \cdot p \lesssim m^2$, the spectrally resolved
cross section $d \sigma /du \vert_{u = const}$ as a function of the laser
intensity parameter $\xi$ displays a pronounced 
$\cap$
%\rotatebox{180}{U}
%\raisebox{\depth}{\rotatebox{180}{U}} 
shape for $u > u_{KN}$ (the ``rise and fall").
This behavior is in stark contrast with the monotonously rising integrated
probability $\lim_{\chi, \xi \to \infty} \mathbb{P} \propto \alpha \chi^{2/3} m^2 /
k \cdot p$. That is, in different regions of the phase space, also different
sensitivities of cross sections/rates/probabilities 
on the laser intensity impact can be observed.
The soft (small-$u$) part, which determines the integrated cross section/probability,
may behave completely different than the hard (large-$u$) contribution.\footnote{An analog
situation is known in QCD \cite{Horowitz:2010yg}: Tree-level diagrams are calculated primarily
with \underline{constant} $\alpha_{QCD}$. Renormalization improvement
means then replacing $\alpha_{QCD}$ by  $\alpha_{QCD} (\mathfrak{s})$ which accounts 
at the same time for all vertices in the considered diagram by one global
scale $\mathfrak{s}$. Inserting explicitly loop corrections leads finally to
scale dependent couplings $\alpha_{QCD} (\mathfrak{s}_v)$ specific for each vertex $v$ separately.} 
Transferred to certain approximations used in simulation codes such 
a behavior implies that one should test differentially where the conditions
for applicability are ensured.   

The hard photons, once produced by Compton process in a laser pulse, 
act as seeds for secondary processes,
most notably the Breit-Wheeler process. A folding model of type
Compton$\otimes$Breit-Wheeler on the probabilistic level
points to a rapidly increasing rate of
$e^+ e^-$ production in the region $\xi  \lesssim 4$, when using
parameters in reach of the planned LUXE set-up. 
The actual plans (see figure 2.10 in LUXE CDR \cite{Abramowicz:2021zja})
uncover $\xi = 2$ (40 TW, 8 $\mu$m laser), 6 (40 TW, 3 $\mu$m) and 16 (300 TW, 3 $\mu$m), and
E-320 envisages $\xi = 10$. The folding model may be utilized as reference to identify the occurrence of the wanted one-step trident process in this energy-intensity regime.
Furthermore, bandwidth effects in the trident process are isolated by considering
the weak-field regime $\xi \to 0$. 

\begin{appendix}

\section{Basics of nonlinear Compton process} \label{App_A}

Following \cite{SeiptDaniel:2012tua} we recall the basics of the underlying
formalism of the nonlinear Compton process. Within the Furry picture the 
lowest-order, tree-level $S$ matrix element for the one-photon 
(four-momentum $k'$, four-polarization $\epsilon'$) decay
of a laser-dressed electron $e_L$ in the background field (\ref{laser})
$e_L (p) \to e_L (p')+ \gamma (k', \epsilon')$ reads
with suitable normalizations of the wave functions
\begin{equation} \label{S1}
S_{fi} = - i e \int d^4 x \, J \cdot {\epsilon^*}' \, %\frac{\exp\{i k' \cdot x \}}{\sqrt{2 \omega'}},
\exp\{i k' \cdot x \}
\end{equation}
where the current $J_\mu (x) = \bar \Psi_{p'} \gamma_\mu \Psi_p $
%\frac{\bar u_{p'} \bar E_{p'}}{\sqrt{2 q_0'}} \gamma_\mu
%\frac{E_p u_p}{\sqrt{2 \q_0}}
%\exp\{ - i (q - q') \cdot x\}
%\exp{\ i e (S_p - S_{p'} \}$ 
is built by the Volkov wave function
$\Psi_p = E_p u_p \exp \{ - i p \cdot x \} \exp\{ - i f_p\}$ 
(spin indices are suppressed) and its adjoint $\bar \Psi$
with Ritus matrix $E_p = 1 + \frac{e}{2 k \cdot p} \slashed k \slashed A$
and phase function
$f_p (\phi) = \frac{e}{k \cdot p}\int_0^\phi d \phi'
\left[p \cdot A - \frac{e}{2} A \cdot A \right]$. We employ Feynman's slash
notation and denote scalar products by the dot between four-vectors;
$u_p$ is the free Dirac bi-spinor.
Exploiting the symmetry of the background field, $A(\phi = k \cdot x)$,
Eq.~(\ref{S1}) can be manipulated (cf.\ \cite{SeiptDaniel:2012tua} for details)
to arrive at
\begin{equation} \label{S2}
S_{fi} = - i e (2 \pi)^3 \frac{2}{k_-} \, \delta^{(3)} 
(\underline{p} - \underline{p'} - \underline{k'}) \,
{\cal M} (\ell) ,
\end{equation}
where $\ell \equiv (k'_- + p'_- - p_-)/ k_- = k' \cdot p / k \cdot p' $
accomplishes the balance equation $p + \ell k - p' - k' = 0$. 
(See \cite{Titov:2015pre} for a formulation with 
$S_{fi} = - i e (2 \pi)^4 \int \frac{d \ell}{2 \pi} \, 
\delta^{(4)} (p + \ell k - p' -k') \, {\cal M} (\ell)$.)
Light-cone coordinates
are useful here, e.g.\ $k_- = k^0 - k^3$, $k_+ = k^0 + k^3$, 
$k_\perp = (k^1, k^2)$, and $\underline k = (k_+, k_\perp)$.
Imposing gauge invariance yields the matrix element
\begin{equation} \label{M1}
{\cal M} = \textstyle \sum_{i=1}^3 J^{(i)} S^{(i)} 
\end{equation}
with the pieces of the electron current
\begin{eqnarray} \label{J1}
\hat J^{(0)} &=& \bar u_{p'} {\slashed \epsilon'}^* u_p, \\
\hat J^{(1, 2)} &=& \bar u_{p'} \left( 
d_{p'} {\slashed \epsilon_{\pm}} \slashed k {\slashed \epsilon'}^* +
d_{p} {\slashed {\epsilon'}}^* \slashed k {\slashed \epsilon_{\pm}} \right) u_p,  \label{J2}\\
\hat J^{(3)} &=& 4 k \cdot {\epsilon'}^* d_{p'} d_p\,  \bar u_{p'} \slashed k u_p, \label{J3}
\end{eqnarray}
which combine to
%\begin{eqnarray}
$J^{(1,2)} = \hat J^{(1,2)} + \frac{\alpha_{\pm}}{2 \ell} \hat J^{(0)}$, %\\
$J^{(3)} = \hat J^{(3)} + \frac{\beta}{\ell} \hat J^{(0)} $.
%\end{eqnarray}
The following abbreviations are used:
$d_p = \frac{m \xi}{4 k \cdot p}$ and $d_{p'} = \frac{m \xi}{4 k \cdot p'}$,
$\alpha_\pm = m \xi \left( \frac{\epsilon_\pm \cdot p}{k \cdot p} - 
\frac{\epsilon_\pm \cdot p'}{k \cdot p'}\right)$ with
$\vec \epsilon_\pm = (\vec a_x \pm i \vec a_y) e^2 / m^2 \xi^2$ and
$\beta = \frac{m^2 \xi^2}{4} \left( \frac{1}{k \cdot p} - 
\frac{1}{k \cdot p'} \right)$ as well.

The phase integrals $S^{(i)}$ are the remainders of the integration 
$d^4 x = d \phi \, dx_- \, d^2 x_\perp / k_-$
in Eq.~(\ref{S1}):
\begin{eqnarray}
S^{(1,2)} &=& \int_{- \infty}^\infty d \phi  \, f(\phi) \,
\exp\{ i (\ell \pm 1) \phi - i (f_p (\phi) - f_{p'} (\phi) ) \}, \nonumber \\
S^{(3)} &=&  \int_{- \infty}^\infty d \phi  \, f^2(\phi)
\exp\{ i \ell \phi - i (f_p (\phi) - f_{p'} (\phi) )\} \label{S(3)}\\
&& \times \left[ 1 + \cos 2 \phi \right]. \nonumber 
\end{eqnarray}
For a few special non-unipolar (plane-wave) fields and their envelopes $f(\phi)$,
the phase integrals can be processed exactly by analytic means 
\cite{Seipt:2016rtk,Dinu:2013hsd},
but in general a numerical evaluation is needed. The very special IPA case of
$f(\phi) = 1$ allows for a simple representation of $f_p(\phi)$ with 
subsequent decomposition of $S^{(i)}$ into Bessel functions, yielding
final expressions as in Eqs.~(\ref{IPA1}, \ref{rate_C_IPA_cont}).
 
For identifying the weak-field limit, $\xi \to 0$, it is necessary to recognize in Eq.~(\ref{M1})
$J^{(0)} \propto \xi^0$, and
$\alpha_\pm, \, J^{(1, 2)} \propto \xi^1$, and
$\beta, \, J^{(3)} \propto \xi^2$.
Thus, $\lim_{\xi \to 0} {\cal M} = {\cal M}_1 \xi +  {\cal M}_2 \xi^2 + \cdots$.
We emphasize the Fourier transform of the pulse envelope, 
$\lim_{\xi \to 0} S^{(1,2)} = \int_{- \infty}^\infty d \phi \, f(\phi) 
\exp\{ i (\ell \pm 1) \phi \}$,
entering ${\cal M}_1$, where only $S^{(1)}$ with $\ell \ge 0$ contributes to
the wanted one-photon emission:\footnote{
The same reasoning applies in \ref{subthreshold} for one-pair emission,
i.e.\ the finite-width Fourier transform of $f(\phi)$ at $\Delta \phi < \infty$
enables the sub-threshold pair production
$e_L(p) \to e_L(p') + e_L(p'') + \bar e_L(p''')$.}
${\cal M}_1 (\ell) = 
J^{(1)}  \int_{- \infty}^\infty d \phi \, f(\phi)  \exp\{ i (\ell - 1) \phi \}$,
$J^{(1)} (\ell) = \frac{m}{2} \epsilon_\mu'{}^* \epsilon_{+ \nu} \bar u_{p'}
\left[ 
\frac{\gamma^\mu \slashed k_\ell \gamma^\nu + 2 \gamma^\mu p^\nu}{2 k_\ell \cdot p}
+ 
\frac{\gamma^\nu \slashed k' \gamma^\mu - 2 \gamma^\nu p^\mu}{2 k' \cdot p}
\right] u_p$ with $k_\ell \equiv \ell k$. For pulses with broad support,
$f (\phi)  \to 1$, within the interval $\Delta \phi \gg 1$, one arrives at the 
standard Compton (Klein-Nishina) expression in leading order by using
$\xi \to 2 e / m$. The appearance
of $\delta (\ell - 1)$ combined with the definition of $\ell$ below Eq.~(\ref{S2})
leads to the famous Compton formula via
$\delta \left( \omega' - \frac{\omega m}{m + \omega (1 - \cos \Theta')} \right)$
in the electron's rest frame by the subsequent phase space integration(s). 

The discussion of the large-$\xi$ limit needs some care in general,
cf.\ \cite{Ilderton:2018nws,DiPiazza:2018bfu}. Useful limits are obtained
for the IPA case \cite{Ritus} under the side condition
$\xi^2 (1 - z^2/n^2) = const$ 
(cf.\ Eqs.~(\ref{rate_C_IPA}, \ref{rate_C_IPA_cont})), e.g.\
%$\overline{\vert {\cal M}^2 \vert}\vert_{u \ll 1}
$\vert {\cal M} \vert^2 \vert_{u \ll 1}
\propto (\xi /u)^{2/3} $ and  
%$ \overline{\vert {\cal M}^2 \vert}\vert_{u \gg 1} 
$\vert {\cal M}^2 \vert_{u \gg 1} 
\propto u^{-1} (\xi / u)^{1/2}
\exp\{ - \frac{2 u}{3 \xi} \frac{m^2}{k \cdot p}\} \to \sqrt{\xi}$
(cf.\ Eqs.~(\ref{region_II}) and (\ref{region_I})), 
At $\xi \gg 1$, the $\xi$ and $u$ dependencies of resulting probabilities for circular polarization 
and constant cross field backgrounds coincide.

These limits are at the heart of the ``rise and fall".

The differential emission probability per $in$-electron and per laser pulse follows from (\ref{S2})
by partial integration over the $out$-phase space,
\begin{equation}
\frac{d  \mathbb{P} }{d \omega'  d \Omega'} = 
\frac{e^2 \omega'}{64 \pi^3 \, k \cdot p \, k \cdot p'}
\vert {\cal M} \vert^2 ,
\end{equation}
and may be transformed to other coordinates, e.g.\
$u$ or $\ell$ etc.
Having in mind the IPA limit, one should turn to the dimensionless differential rate \cite{Titov:2015pre}
\begin{equation}
\frac{d  \Gamma_C}{d \omega'  d \Omega'} = 
\frac{e^2 \omega'}{32 \pi^2 \, q_0 \, k \cdot p'}
\vert {\cal M} \vert^2 .
\end{equation}
Spin averaging of the $in$-electron, and spin summation of the
$out$-electron and summation over the $out$-photon polarizations leads to
$\overline{\vert {\cal M} \vert^2}$,
unless one is interested in polarization effects as in 
\cite{Seipt:2020diz,King:2020btz,Ivanov:2004vh}.
The above expressions (\ref{M1} - \ref{S(3)}) can be further processed
for special field envelopes, or $\vert {\cal M} \vert^2$ is numerically accessible, 
via Eq.~(\ref{M1}), as mod-squared sum of complex number products
provided by Eqs.~(\ref{J1} - \ref{S(3)})
which need afterwards explicit (numerical) spin and polarization summation/averaging to arrive at $\overline{\vert {\cal M} \vert^2}$.

The cross section is obtained 
by normalization on the integrated laser photon flux:
$d \sigma = d  \Gamma_C \frac{q_0}{k \cdot p} \frac{\omega}{n_L}$ 
with $n_L = \frac{m^2}{e^2} \xi^2 \omega N_L$, where
$N_L = 1$ (IPA, quasi-momentum $q_0$) or 
$N_L = \frac{1}{2 \pi} \int_{- \infty}^\infty d \phi \, f(\phi)$ (FPA, $q_0 \equiv p_0$).
The circularly polarized laser background (\ref{laser}) is supposed in these relations.

% The treatment of the Breit-Wheeler process as cross channel goes analog.
\end{appendix}

\vspace*{-8mm}

\begin{acknowledgments}
%{\bf Acknowledgments:}
The authors gratefully acknowledge the collaboration with 
D.~Seipt, T.~Nousch, T.~Heinzl, %U.~Hernandez Acosta 
and useful discussions with
A.~Ilderton, K.~Krajewska,  M.~Marklund, C.~M\"uller, S.~Rykovanov, 
and G.~Torgrimsson.
A.~Ringwald and B.~King are thanked for explanations w.r.t.\ LUXE.
The work is supported by R.~Sauerbrey and T.~E.~Cowan w.r.t.\ the study
of fundamental QED processes for HIBEF.
\end{acknowledgments}

\end{document}